\documentclass[5p,times]{elsarticle}
\usepackage{mathrsfs,bm}
\usepackage{longtable,lscape}
\usepackage{txfonts}
\usepackage{indentfirst}
\usepackage{graphicx,color,dcolumn,booktabs}
\usepackage{multirow}
\usepackage{indentfirst}
\usepackage{graphicx,color,dcolumn,booktabs}

\usepackage{amsmath}

\usepackage{color}

\definecolor{cover}{rgb}{0.77,0.87,0.88}
\definecolor{blueone}{rgb}{0.1,0.1,.7}
\definecolor{citec}{rgb}{0.14,0.47,0.09}
\definecolor{two}{rgb}{0.0,0.5,0.}
\definecolor{three}{rgb}{.5,.1,0.15}
\usepackage[bookmarks=true,bookmarksopen=false,plainpages=false,breaklinks=true,
   bookmarksnumbered=true,hypertexnames=false,
   filecolor=blue,urlcolor=three,menucolor=three,
   linkcolor=three,citecolor=blueone, colorlinks,
   anchorcolor=blue,runcolor=pink,frenchlinks=red
   pdfstartview=FitH,pdftitle=title,%
   pdfauthor=author]{hyperref}

\allowdisplaybreaks[4]

\begin{document}

\begin{frontmatter}
\title{Study of $\Lambda p$ and $\bar{\Lambda} p$ scatterings via quasipotential Bethe-Salpeter equation}
\author{Yaning Chen}
\author{Jun He}\ead{Corresponding author: junhe@njnu.edu.cn}
\address{School of Physics and Technology, Nanjing
Normal University, Nanjing 210097, China}

\begin{abstract}

Motivated by recent BESIII measurements of the $\Lambda p \to \Lambda p$ and $\bar{\Lambda} p \to \bar{\Lambda} p$ scattering processes, we investigate these reactions within the framework of the quasipotential Bethe-Salpeter equation using an effective Lagrangian approach. The interaction potentials are constructed via a one-boson-exchange model incorporating pseudoscalar, scalar, and vector meson exchanges, along with coupled-channel effects from the $\Sigma N$ and $\bar{\Sigma} N$ channels.
For the $\Lambda p \to \Lambda p$ reaction, the total cross sections from threshold up to $\sqrt{s} = 2.5~\text{GeV}$ are well reproduced. A mild enhancement near the $\Sigma N$ threshold is attributed to coupled-channel dynamics. Using parameters constrained by the total cross section data, our model also predicts differential cross sections at $\sqrt{s} = 2.24~\text{GeV}$, which exhibit weak angular dependence, consistent with experimental observations. Partial-wave analysis indicates that the $1^+$ partial wave dominates over the entire energy range, while the $0^+$ wave plays a significant role near threshold.
For the $\bar{\Lambda} p \to \bar{\Lambda} p$ reaction, our predicted total cross sections show good agreement with the BESIII data. The $1^-$ partial wave is found to dominate in most of the energy region. Notably, the calculated differential cross sections exhibit a strong forward peaking behavior, consistent with experimental findings and understood as resulting from constructive interference among various partial waves. This forward-peaked angular distribution persists across a range of energies, highlighting the distinct dynamics of the $\bar{\Lambda} p$ interaction.
\end{abstract}

  \end{frontmatter}
\section{INTRODUCTION}

In nuclear physics, hyperon-nucleon ($YN$) interactions represent a crucial
extension of the study of baryon-baryon interactions. While nucleon–nucleon
($NN$) interactions are relatively well understood~\cite{Machleidt:1987hj}, the
$YN$ system remains poorly constrained due to experimental limitations,
hindering the development of a unified theoretical
framework~\cite{ParticleDataGroup:2022pth}. This deficiency has far-reaching
implications, particularly for the physics of neutron
stars~\cite{Tolos:2020aln,Vidana:2018bdi,Vidana:2013nxa}. Hyperons are expected
to emerge in the dense cores of neutron stars, and their presence can
significantly influence stellar structure. However, the appearance of hyperons
tends to soften the equation of state (EOS), making it difficult to reconcile with the
observed existence of massive neutron stars—a challenge commonly referred to as
the “hyperon puzzle”~\cite{Tolos:2020aln,Vidana:2018bdi,Vidana:2013nxa}.
Addressing this problem requires a more precise understanding of $YN$
interaction mechanism in order to improve the modeling of the EOS via realistic
interaction potentials.  Among various approaches, the most direct method to
investigate $YN$ interactions is through scattering
experiments~\cite{Alexander:1968acu,Kadyk:1971tc,Sechi-Zorn:1968mao}. However,
such studies face serious experimental challenges, including the instability of
hyperon beams and the short lifetimes of hyperons. As a result, available
scattering data are scarce, leading to large uncertainties in existing
interaction models.

The $\Lambda p \to \Lambda p$ scattering process represents the most extensively studied channel in $YN$ interactions~\cite{ParticleDataGroup:2022pth}. Early experimental measurements date back to the 1960s and 1970s~\cite{Alexander:1968acu,Kadyk:1971tc,Sechi-Zorn:1968mao,HAUPTMAN197729,Herndon:1967zz,PhysRevLett.2.174,Roe:1961zz,PhysRev.129.1372,BEILLIERE1964350,PhysRevLett.12.625,PhysRevLett.13.282,Bassano:1967kbh,Charlton:1970bv,Anderson:1975rh,Mount:1975pb}, during which a range of investigations covered both low- and high-energy regimes. However, these studies were significantly limited by the instability of hyperon beams and the restricted resolution of bubble chamber detectors~\cite{DODD1955}, leading to a stagnation in the field for several decades.
This situation changed in 2021, when the CLAS Collaboration reported high-precision measurements of the $\Lambda p \to \Lambda p$ cross sections using modern detector technologies~\cite{CLAS:2021gur}, marking a major experimental advance and offering valuable constraints for theoretical models.

On the theoretical side, $\Lambda p \to \Lambda p$ scattering has been explored using various approaches. The constituent quark model, originally developed for $NN$ interactions, has been extended to $YN$ systems~\cite{Zhang:1997ny,Fujiwara:2006yh}. Meson-exchange models have also been widely employed, notably by the Jülich and Nijmegen groups~\cite{Rijken:1998yy,Holzenkamp:1989tq,Reuber:1992dh,Haidenbauer:2005zh,Nagels:1977ze,Maessen:1989sx,Rijken:2010zzb,Tominaga:2001ra}, to construct detailed descriptions of $YN$ scattering. More recently, chiral effective field theory has enabled systematic investigations at leading order (LO) and next-to-leading order (NLO)~\cite{Haidenbauer:2019boi,Song:2021yab,Polinder:2006zh,Haidenbauer:2013oca,Li:2016paq,Li:2016mln,Haidenbauer:2023qhf}. Despite these theoretical developments, further experimental input remains essential to constrain model parameters and reduce uncertainties.

A major breakthrough was recently achieved by the BESIII Collaboration~\cite{BESIII:2024geh}, which reported measurements of $\Lambda p \to \Lambda p$ scattering, including differential cross sections that are rarely available in the literature. More notably, BESIII also performed the first measurement of differential cross sections for $\bar{\Lambda}p \to \bar{\Lambda}p$, marking the beginning of experimental studies on antihyperon-nucleon interactions. Following this, Wang et al.~\cite{Wang:2024whi} provided a dynamical interpretation of the observed total and differential cross sections for both processes using tree-level Feynman diagrams. While their results show good agreement with the BESIII data, the analysis was limited to the specific energy point measured by BESIII and did not account for rescattering effects.

Motivated by these experimental and theoretical developments, we carry out a unified analysis of the $\Lambda p \to \Lambda p$ and $\bar{\Lambda}p \to \bar{\Lambda}p$ scattering processes within the quasi-potential Bethe-Salpeter equation (qBSE) framework, employing an effective Lagrangian approach. The interaction potentials are derived from a one-boson-exchange model incorporating pseudoscalar, scalar, and vector meson exchanges. A partial wave decomposition is performed at the amplitude level to obtain the scattering amplitudes for individual partial waves, which are then summed to calculate the total and differential cross sections over the energy range from threshold to $\sqrt{s}=2.5$~GeV. The results are systematically compared with available experimental and theoretical studies, and angular distributions at selected energies are also presented.

The structure of this paper is as follows. In Section~\ref{Sec: Formalism}, we present the theoretical formalism, including the qBSE framework and interaction potentials. Numerical results and comparisons are discussed in Section~\ref{sec3}. Finally, conclusions and implications are summarized in Section~\ref{sum}.

\section{Formalism}\label{Sec: Formalism}

\begin{figure}[h!]
  \centering
  \includegraphics[
    width=0.48\textwidth,  
    trim=0 480 0 10,     
    clip                  
    ]{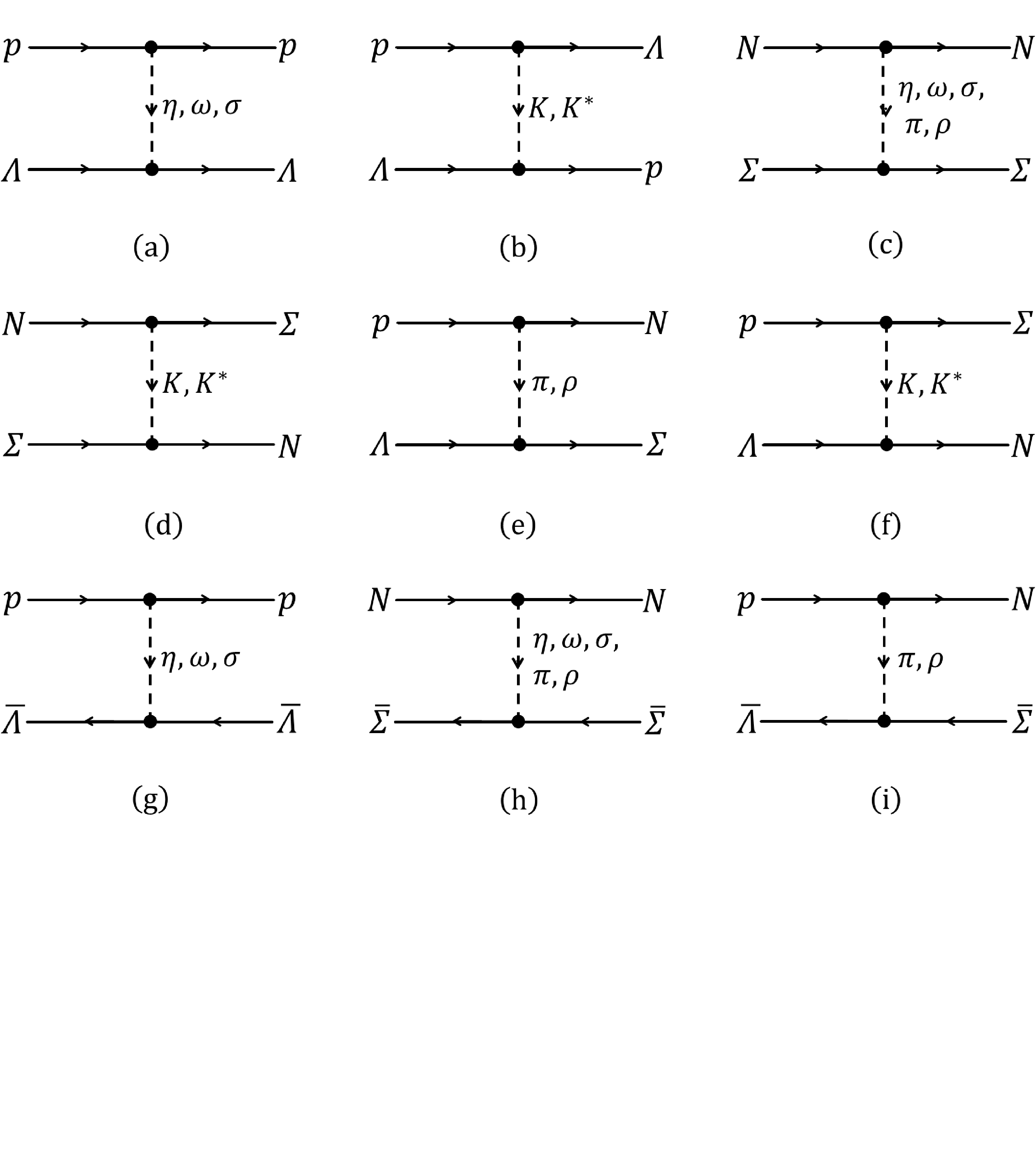}
  \caption{Feynman diagrams illustrating the interactions included in the present study. Diagrams (a)–(f) represent processes involved in $\Lambda p$ scattering, whereas diagrams (g)–(i) correspond to $\bar{\Lambda} p$ scattering.}
  \label{feyn}
\end{figure}

In the present work, we focus on the scattering processes $\Lambda p \to \Lambda p$ and $\bar{\Lambda} p \to \bar{\Lambda} p$. For the former, the direct process occurs via the $t$-channel through $\eta$, $\omega$, and $\sigma$ exchanges, and via the $u$-channel through $K$ and $K^*$ exchanges, as illustrated in Fig.~\ref{feyn}(a) and Fig.~\ref{feyn}(b), respectively. For the latter process, only the $t$-channel contribution is allowed, as shown in Fig.~\ref{feyn}(g).

Owing to the small mass difference of approximately 77 MeV between the $\Sigma$ and $\Lambda$ hyperons, the $\Sigma N$ channel lies in the energy region of interest in this work. Thus, the coupled-channel effects involving this channel should not be neglected. We therefore incorporate the coupling to the $\Sigma N$ channel, following Ref.~\cite{Holzenkamp:1989tq}. Specifically, the $\Sigma N \to \Sigma N$ process is included, as shown in Fig.~\ref{feyn}(c) and Fig.~\ref{feyn}(d), along with the transition between the $\Lambda p$ and $\Sigma N$ channels, illustrated in Fig.~\ref{feyn}(e) and Fig.~\ref{feyn}(f). Similarly, for the $\bar{\Lambda} p$ scattering, the $\bar{\Sigma} N$ interaction and its coupling to the $\bar{\Lambda} p$ channel are taken into account, as depicted in Fig.~\ref{feyn}(h) and Fig.~\ref{feyn}(i).

In all cases, we adopt the isospin basis but construct the $\Sigma N$ channel by explicitly requiring the total electric charge to be positive, consistent with the $\Lambda p$ system. Specifically, the allowed $\Sigma N$ combinations form the isospin-$\frac{1}{2}$ state with $I_3 = +\frac{1}{2}$ as
\begin{equation}
|I=1/2, I_3=+1/2\rangle = -\sqrt{\frac{2}{3}}|\Sigma^+ n\rangle - \sqrt{\frac{1}{3}}|\Sigma^0 p\rangle,
\end{equation}
and the $\bar{\Sigma} N$ channel is constructed analogously.

To describe the interactions, we need the vertices for the baryons and the exchanged mesons. The interaction amplitudes can be constructed using standard Feynman rules, based on the effective Lagrangians corresponding to these vertices. In this work, the relevant Lagrangians are formulated using effective field theory approaches incorporating SU(3) flavor symmetry and chiral symmetry as~\cite{Ronchen:2012eg,Kamano:2008gr,Kong:2023dwz},
\begin{align}
\mathcal{L}_{NN\pi}
&= -\frac{g_{NN\pi}}{m_{\pi}} \bar{N} \gamma^5 \gamma^{\mu} \bm{\tau} \cdot \partial_{\mu} \bm{\pi} N, \nonumber\\
\mathcal{L}_{NN\eta}
&= -\frac{g_{NN\eta}}{m_{\pi}} \bar{N} \gamma^5 \gamma^{\mu} \partial_{\mu} \eta N, \nonumber\\
\mathcal{L}_{NN\sigma}
&= -g_{NN\sigma} \bar{N} N \sigma.\nonumber\\
\mathcal{L}_{NN\rho} &=
-g_{NN\rho}\bar{N}[\gamma^{\mu}-\frac{\kappa_{NN\rho}}{2m_{N}}\sigma^{\mu\nu}\partial_{\nu}]
\bm{\tau}\cdot\bm{\rho_{\mu}}N,
\nonumber\\
\mathcal{L}_{NN\omega}
&=-g_{NN\omega}\bar{N}[\gamma^{\mu}-\frac{\kappa_{NN\omega}}{2m_{N}}\sigma^{\mu\nu}\partial_{\nu}]
\omega_{\mu}N,
\nonumber\\
\mathcal{L}_{\Lambda \Lambda \eta}
&= -\frac{g_{\Lambda \Lambda \eta}}{m_{\pi}} \bar{ \Lambda } \gamma^5 \gamma^{\mu}  \partial_{\mu} \eta \Lambda  , \nonumber\\
\mathcal{L}_{\Lambda \Lambda \sigma}
&= -g_{\Lambda \Lambda \sigma} \bar{\Lambda  }  \Lambda  \sigma.\nonumber\\
\mathcal{L}_{\Lambda \Lambda \omega}
&=-g_{\Lambda \Lambda \omega}\bar{\Lambda  }[\gamma^{\mu}-\frac{\kappa_{\Lambda \Lambda \omega}}{2m_{N}}\sigma^{\mu\nu}\partial_{\nu}]
\omega_{\mu}\Lambda,
\nonumber\\
\mathcal{L}_{\Sigma\Sigma\pi}
&= i\frac{g_{\Sigma\Sigma\pi}}{m_{\pi}} \bm{{\bar\Sigma}}\times\gamma^5 \gamma^{\mu} \bm{\Sigma} \cdot  \partial_{\mu} \bm{\pi} , \nonumber\\
\mathcal{L}_{\Sigma\Sigma\eta}
&= -\frac{g_{\Sigma\Sigma\eta}}{m_{\pi}} \bm{\bar{\Sigma}}\cdot\gamma^5 \gamma^{\mu} \bm{\Sigma} ~\partial_{\mu} \eta , \nonumber\\
\mathcal{L}_{\Sigma\Sigma\sigma}
&= -g_{\Sigma\Sigma\sigma} \bm{\bar{\Sigma}} \cdot\bm{\Sigma} \sigma , \nonumber\\
\mathcal{L}_{\Sigma\Sigma\rho} &=
ig_{\Sigma\Sigma\rho}\bm{\bar{\Sigma}}\times[\gamma^{\mu}-\frac{\kappa_{\Sigma\Sigma\rho}}{2m_{N}}\sigma^{\mu\nu}\partial_{\nu}]\bm{\Sigma}
\cdot\bm{\rho_{\mu}},
\nonumber\\
\mathcal{L}_{\Sigma\Sigma\omega}
&=-g_{\Sigma\Sigma\omega}\bm{\bar{\Sigma}}\cdot[\gamma^{\mu}-\frac{\kappa_{NN\omega}}{2m_{N}}\sigma^{\mu\nu}\partial_{\nu}]\bm{\Sigma}
~\omega_{\mu},
\nonumber\\
\mathcal{L}_{N\Lambda K}
&= -\frac{g_{N\Lambda K}}{m_{\pi}} \bar{N} \gamma^5 \gamma^{\mu} \Lambda ~\partial_{\mu} K, \nonumber\\
\mathcal{L}_{N\Sigma K}
&= -\frac{g_{N\Sigma K}}{m_{\pi}} \bar{N} \gamma^5 \gamma^{\mu} \bm{\tau} \cdot  \bm{\Sigma}~\partial_{\mu}  K, \nonumber\\
\mathcal{L}_{N\Lambda K^*} &=
-g_{N\Lambda K^*}\bar{N}[\gamma^{\mu}-\frac{\kappa_{N\Lambda K^*}}{2m_{N}}\sigma^{\mu\nu}\partial_{\nu}]\Lambda~K^*_{\mu},
\nonumber\\
\mathcal{L}_{N\Sigma K^*} &=
-g_{N\Sigma K^*}\bar{N}[\gamma^{\mu}-\frac{\kappa_{N\Sigma K^*}}{2m_{N}}\sigma^{\mu\nu}\partial_{\nu}]\bm{\tau}\cdot\bm{\Sigma}~
K^*_{\mu},
\nonumber\\
\mathcal{L}_{\Lambda\Sigma\pi}
&= -\frac{g_{\Lambda\Sigma\pi}}{m_{\pi}} \bar{\Lambda} \gamma^5 \gamma^{\mu}\bm{\Sigma} \cdot\partial_{\mu} \bm{\pi}, \nonumber\\
\mathcal{L}_{\Lambda\Sigma\rho} &=
-g_{\Lambda\Sigma\rho}\bar{\Lambda}[\gamma^{\mu}-\frac{\kappa_{\Lambda\Sigma\rho}}{2m_{N}}\sigma^{\mu\nu}\partial_{\nu}]\bm{\Sigma}\cdot
\bm{\rho_{\mu}},
,\label{LD} \end{align}
Here, $\pi$, $\eta$, $\sigma$, $\omega$, $\rho$, $K$, and $K^*$ denote mesonic fields; $\Lambda$ and $\Sigma$ correspond to hyperonic fields; and $N$ represents the nucleonic field. The particle masses employed in the present work are taken from the central values recommended by the Particle Data Group (PDG)~\cite{ParticleDataGroup:2018ovx}, with the $\sigma$ meson mass fixed at $550~\text{MeV}$. Although a variety of effective Lagrangians are introduced for different baryon-meson interactions, they involve only three distinct Lorentz structures, which are further combined with isospin structures based on SU(3) flavor symmetry. Accordingly, the coupling constants associated with the vertices involving pseudoscalar and vector mesons are determined via SU(3) flavor symmetry relations~\cite{Ronchen:2012eg}. The numerical values adopted in this work are summarized in Table~\ref{coupling}.
\renewcommand\tabcolsep{0.295cm}
\renewcommand{\arraystretch}{1.}
\begin{table}[h!]
\caption{Coupling constants determined using SU(3) flavor symmetry relations, as adopted in our calculation. All values are dimensionless (in units of 1), and taken from Ref.~\cite{Ronchen:2012eg}. \label{coupling}}
\begin{tabular}{cccccc}\toprule[1pt]
$g_{NN\pi}$&$g_{NN\eta}$&$g_{NN\omega}$&$\kappa_{NN\omega}$&$g_{NN\rho}$&$\kappa_{NN\rho}$\\
$0.989$&$0.147$&$9.75$&$0$&$3.25$&$19.82$\\\hline
$g_{\Lambda \Lambda \eta}$&$g_{\Lambda \Lambda \omega}$&$\kappa_{\Lambda\Lambda\omega}$&$g_{\Sigma\Sigma\pi}$&$g_{\Sigma\Sigma\eta}$\\
$-0.682$&$6.5$&$-9.91$&$0.791$&$0.682$\\\hline
$g_{\Sigma\Sigma\omega}$&$\kappa_{\Sigma\Sigma\omega}$&$g_{\Sigma\Sigma\rho}$&$\kappa_{\Sigma\Sigma\rho}$&$g_{\Lambda\Sigma\pi}$&$g_{\Lambda\Sigma\rho}$\\
$6.5$&$9.91$ &$6.5$&$9.91$ &$0.682$&$0$\\\hline
$g_{N\Lambda K}$&$g_{N\Sigma K}$&$g_{N\Lambda K^*}$&$\kappa_{N\Lambda K^*}$&$g_{N\Sigma K^*}$&$\kappa_{N\Sigma K^*}$\\
$-1.03$&$0.198$&$-5.63$&$-17.20$&$-3.25$&$9.91$\\
\bottomrule[1pt]
\end{tabular}
\end{table}

The coupling constants for the scalar $\sigma$ meson cannot be determined from SU(3) symmetry alone, and there remains significant uncertainty regarding the strength of $\sigma$ exchange. In this study, we adopt $g_{\sigma NN} = 9.42$ from the Bonn nucleon-nucleon potential~\cite{Machleidt:1987hj}, and $g_{\Sigma\Sigma\sigma} = 3.1152$ from the Ehime OBEP framework~\cite{Tominaga:2001ra}. The coupling constant $g_{\Lambda\Lambda\sigma}$ is treated as a free parameter, and its specific values are discussed in Section~\ref{sec3}.

With the effective Lagrangians above, the vertex structures $\Gamma_{1,2}$ for the upper and lower interaction vertices can be constructed. Together with the meson propagators $P$, the interaction potentials are derived using standard Feynman rules. The resulting potentials take the following form, following the approach in Ref.~\cite{He:2019rva}:
\begin{equation}%
{\cal V}_{\mathbb{P},\sigma}=f_I\Gamma_1\Gamma_2 P_{\mathbb{P},\sigma}f(q^2),\ \
{\cal V}_{\mathbb{V}}=f_I\Gamma_{1\mu}\Gamma_{2\nu}  P^{\mu\nu}_{\mathbb{V}}f(q^2).\label{V}
\end{equation}
The propagators are defined as 
\begin{equation}%
P_{\mathbb{P},\sigma}= \frac{i}{q^2-m_{\mathbb{P},\sigma}^2},\ \
P^{\mu\nu}_\mathbb{V}=i\frac{-g^{\mu\nu}+q^\mu q^\nu/m^2_{\mathbb{V}}}{q^2-m_\mathbb{V}^2}.
\end{equation}
We introduce a form factor $f(q^2) = e^{-(m_e^2 - q^2)^2 / \Lambda_e^2}$ to account for the off-shell effect of the exchanged meson, where $m_e$ and $q$ denote the mass and momentum of the exchanged meson, respectively. The parameter $\Lambda_e$ serves as a cutoff to suppress contributions from highly off-shell regions.
To eliminate unphysical singularities in the meson propagator, we follow the prescription of Ref.~\cite{Gross:2008ps} and replace $q^2$ with $-|\vec{q},|^2$.
The factor $f_I$ represents the flavor coefficient associated with a specific meson exchange in a given interaction channel, with its values listed in Table~\ref{flavor factor}.
\renewcommand\tabcolsep{0.23cm}
\renewcommand{\arraystretch}{1.1}
\begin{table}[h!]
\caption{The flavor factors $f_I$ for certain meson exchanges of certain interaction.
 \label{flavor factor}}
 \begin{tabular}{c|ccccccc}\bottomrule[1pt]
  Interaction &$\pi$&$\eta$& $\omega$ &$\rho$&$\sigma$&$K$&$K^*$  \\\hline
 ${\Lambda p \to \Lambda p}$  &0&1&1&0&1&1&1\\
 ${\Lambda p \to \Sigma N}$  &$-\sqrt{3}$&$0$&$0$&$-\sqrt{3}$&$0$&$-\sqrt{3}$&$-\sqrt{3}$ \\
 ${N\Sigma  \to N\Sigma }$  &-2&1&1&-2&1&-1&-1\\
 \toprule[1pt]
 \end{tabular}
\end{table}

In scalar meson exchange interactions, the exchanged mesons typically generate attractive forces at intermediate to long ranges. However, they may also lead to excessive attraction at short distances, potentially resulting in unphysical bound states. To mitigate this short-range overattraction, we introduce a phenomenological repulsive potential, following the methodology of the Nijmegen soft-core model (ESC)~\cite{Rijken:2010zzb}. This repulsive term suppresses contributions from high momentum transfer and is defined as:
\begin{equation}
V_{\text{rep}} = -g_{\text{rep}} \Gamma_1 \Gamma_2,
\end{equation}
where $g_{\text{rep}}$ characterizes the strength of the repulsive interaction. The value of $g_{\text{rep}}$ will be discussed in the following section. 

In this work, we further investigate $\bar{\Lambda}p$ scattering, which involves coupled $\bar{\Lambda}p$ and $\bar{\Sigma}N$ channels as illustrated in Fig.~\ref{feyn} (g-i). The interactions governing these processes can be derived using the G-parity rule~\cite{PHILLIPS:1967wls,Klempt:2002ap,Zhu:2021lhd,Song:2022yfr}, leading to the following relation:
\begin{equation}
\mathcal{V}_{{B}\bar{B}M} = -\mathcal{V}_{BB\pi} + \mathcal{V}_{BB\eta} + \mathcal{V}_{BB\rho} - \mathcal{V}_{BB\omega} + \mathcal{V}_{BB\sigma},
\end{equation}
where the signs on the right-hand side are determined by the G-parity of the exchanged meson $M$. This approach provides a systematic way to connect baryon-antibaryon interactions with the well-established baryon-baryon interaction framework.

The scattering amplitude is obtained by introducing the potential kernel into the Bethe-Salpeter equation. By applying partial wave decomposition and adopting the spectator quasipotential approximation, the equation can be simplified under a fixed spin-parity $J^P$. This procedure reduces the original four-dimensional integral equation in Minkowski space to a one-dimensional integral equation, allowing the scattering amplitude to be calculated efficiently, as detailed in Refs.~\cite{He:2015mja,He:2014nya,He:2012zd,He:2015cea},
\begin{align}
i{\cal M}^{J^P}_{\lambda'\lambda}({\rm p}',{\rm p})
&=i{\cal V}^{J^P}_{\lambda',\lambda}({\rm p}',{\rm
p})+\sum_{\lambda''}\int\frac{{\rm
p}''^2d{\rm p}''}{(2\pi)^3}\nonumber\\
&\cdot
i{\cal V}^{J^P}_{\lambda'\lambda''}({\rm p}',{\rm p}'')
G_0({\rm p}'')i{\cal M}^{J^P}_{\lambda''\lambda}({\rm p}'',{\rm
p}),\quad\quad \label{Eq: BS_PWA}
\end{align}
where, ${\cal M}^{J^P}_{\lambda'\lambda}({\rm p}',{\rm p})$ denotes the partial
wave scattering amplitude, and the propagator $G_0({\rm p}'')$ is simplified
from its original four-dimensional form under the quasipotential approximation.
It takes the form 
\begin{align}G_0({\rm p}'') = \frac{1}{2E_h({\rm p}'')[(W - E_h({\rm p}''))^2 - E_l^2({\rm p}'')]}.
\end{align}
Following the spectator approximation adopted in this work, the heavier particle
(denoted as $h$) in a given channel is placed on shell~\cite{Gross:1991pm},
satisfying $p_h''^0 = E_h({\rm p}'')$. The energy
of the lighter particle (denoted as $l$) is then determined by $p_l''^0 = W -
E_h({\rm p}'')$, where $W$ is the total energy in the center-of-mass frame.
Here, $E_{h,l}({\rm p}'') = \sqrt{m_{h,l}^2 + {\rm p}''^2}$, and  $m_{h,l}$ represent
the  masses of the heavy and light constituent particles, respectively. We define the magnitude of the three-momentum $|{\bm p}|$ as ${\rm p}$. The
partial wave potential is given by:

\begin{align}
{\cal V}_{\lambda'\lambda}^{J^P}({\rm p}',{\rm p})
&=2\pi\int d\cos\theta
~[d^{J}_{\lambda\lambda'}(\theta)
{\cal V}_{\lambda'\lambda}({\bm p}',{\bm p})\nonumber\\
&+\eta d^{J}_{-\lambda\lambda'}(\theta)
{\cal V}_{\lambda'-\lambda}({\bm p}',{\bm p})],
\end{align}
where $\eta = PP_1P_2(-1)^{J-J_1-J_2}$, with $P$ and $J$ representing the system's parity and spin, respectively, alongside those for constituent particles 1 and 2. The initial and final relative momenta are defined as ${\bm p}=(0,0,{\rm p})$ and ${\bm p}'=({\rm p}'\sin\theta,0,{\rm p}'\cos\theta)$. The Wigner d-matrix is expressed as $d^J_{\lambda\lambda'}(\theta)$.

To solve the integral equation in Eq.~(\ref{Eq: BS_PWA}), the momenta ${\rm p}$,
${\rm p}'$, and ${\rm p}''$ are discretized using the Gauss quadrature method
with weights $w({\rm p}_i)$. The discretized form of the 
qBSE can then be written as~\cite{He:2015mja}:
\begin{eqnarray}
{M}_{ik}
&=&{V}_{ik}+\sum_{j=0}^N{ V}_{ij}G_j{M}_{jk}.\label{Eq: matrix}
\end{eqnarray}
The discretized propagator $G_j$ takes the form:
\begin{eqnarray}
	G_{j>0}&=&\frac{w({\rm p}''_j){\rm p}''^2_j}{(2\pi)^3}G_0({\rm
	p}''_j), \nonumber\\
G_{j=0}&=&-\frac{i{\rm p}''_{\rm o}}{32\pi^2 W}+\sum_j
\left[\frac{w({\rm p}_j)}{(2\pi)^3}\frac{ {\rm p}''^2_o}
{2W{({\rm p}''^2_j-{\rm p}''^2_{\rm o})}}\right].
\end{eqnarray}
Here, ${\rm p}''_{\rm o}$ is the on-shell momentum, defined as ${\rm p}''_{\rm o} = \lambda^{1/2}(W, M_1, M_2)/2W$, with the Källén function $\lambda(x,y,z) = [x^2 - (y+z)^2][x^2 - (y-z)^2]$, and $W$ denotes the total energy of the two-body system.
To regularize the propagator and suppress high-momentum contributions, an exponential form factor is introduced:
\begin{equation}
G_0({\rm p}'') \to G_0({\rm p}'')\left[e^{-(p_l''^2 - m_l^2)^2/\Lambda_r^4}\right]^2,
\end{equation}
where $\Lambda_r$ is a cutoff parameter~\cite{He:2015mja}. In this framework, all cutoffs appearing in the form factors—including those in the propagator and in the meson-exchange interactions—are treated as free parameters, and for simplicity, we adopt a common cutoff $\Lambda_e = \Lambda_r = \Lambda$.

The differential cross section is given by
\begin{eqnarray}
\frac{d\sigma}{d\Omega} = \frac{1}{(2j_1+1)(2j_2+1)} \frac{1}{64\pi^2 s} \frac{{\rm p}'}{{\rm p}} \sum_{\lambda', \lambda} \left| M^{J^P}_{\lambda'\lambda}({\rm p}', {\rm p}) \right|^2,
\end{eqnarray}
Here, $s$ is the square of the total energy in the center-of-mass frame. $j_1$ and $j_2$ are the spin of the intitial particles. 

\section{Numerical Result}\label{sec3}
\subsection{Scattering $\Lambda p\to\Lambda p$}

\begin{figure}[h!]
  \centering
  \includegraphics[
    width=0.48\textwidth,  
    trim=0 0 0 0,     
    clip                  
    ]{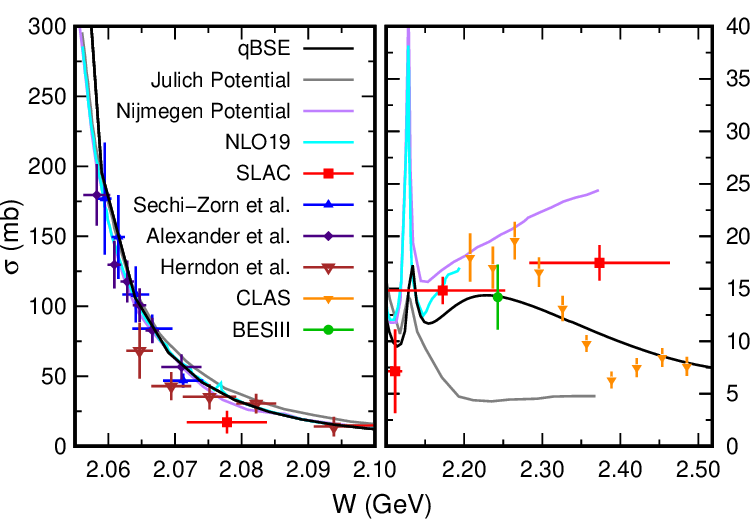}

\caption{Total cross section for the reaction $\Lambda p \to \Lambda p$. The
black solid curve denotes the result obtained using the qBSE approach. Results
for energies below and above 2.1~GeV are presented in the left and right panels,
respectively. For comparison, theoretical predictions are shown from the J\"ulich
potential~\cite{Haidenbauer:2005zh} (grey  curve), the Nijmegen
potential~\cite{Rijken:1998yy} (purple  curve), and
NLO19~\cite{Haidenbauer:2019boi} (cyan  curve). Experimental data are taken from
SLAC~\cite{HAUPTMAN197729} (red  squares), Sechi-Zorn et
al.~\cite{Sechi-Zorn:1968mao} (blue  triangles), Alexander et
al.~\cite{Alexander:1968acu} (indigo  diamonds), Herndon et
al.~\cite{Herndon:1967zz} (brown  inverted triangles), CLAS~\cite{CLAS:2021gur}
(orange  inverted triangles), and BESIII~\cite{BESIII:2024geh} (green
circles).}

\label{fig:cs1}

\end{figure}

Before analyzing the BESIII data at $\sqrt{s} = 2.24$ GeV, we first calculate the total cross sections of the $\Lambda p \to \Lambda p$ process in the energy range from threshold to $\sqrt{s}=2.5$ GeV to provide a global fit to the available experimental data, as shown in Fig.~\ref{fig:cs1}. The fit yields a cutoff parameter $\Lambda = 0.56$ GeV and a coupling constant $g_{\text{rep}} = 11$ for the short-range repulsive interaction, which dominates in the high-energy region. The $\Lambda\Lambda\sigma$ coupling constant is determined to be $g_{\Lambda\Lambda\sigma} = 5$, a value consistent with those reported in the literature, such as 5.79, 6.54, 6.59, 7.58, and 8.17~\cite{Holzenkamp:1989tq,Reuber:1992dh,Zhu:2019ibc}.
In our calculation, we find that the results converge for total angular momentum up to $J \leq 4$. As shown in Fig.~\ref{fig:cs1}, the total cross sections are computed from threshold up to $\sqrt{s} = 2.5$ GeV and are compared with both experimental data and theoretical results from previous studies. For a clearer presentation, we divide the energy range into two regions: from threshold to 2.1 GeV, and from 2.1 to 2.5 GeV. Different vertical axis scales are used in these two regions to improve the visibility of the cross section behavior.

In the lower energy region, from threshold to 2.1 GeV, a large amount of experimental data is available, showing a monotonically decreasing trend. Our results exhibit remarkable consistency with these data. Moreover, the theoretical predictions from the literature, as well as our own calculations, show good agreement with each other in this region.

In the higher energy region above 2.1 GeV, discrepancies between theoretical predictions and experimental data become more pronounced. For instance, the Nijmegen potential reproduces the data around 2.2–2.3 GeV well, but shows a continuously increasing trend~\cite{Rijken:1998yy}, while the J\"ulich potential predicts a smaller and flatter cross section~\cite{Haidenbauer:2005zh}. The CLAS data exhibit a clear enhancement in the cross section around 2.25 GeV~\cite{CLAS:2021gur}, which is also reproduced in our calculation, though the magnitude is somewhat smaller. Our result in this region is also consistent with the recent BESIII data~\cite{BESIII:2024geh}.

Since the $\Sigma N$ threshold lies within the energy range considered in this work, we include the coupling to the $\Sigma N$ channel in our calculations. In the literature, the effects of the $\Sigma N$ channel have also been investigated. For example, both the Nijmegen potential and the NLO19 model predict a pronounced sharp peak near the threshold, as shown in the right panel of Fig.~\ref{fig:cs1}~\cite{Rijken:1998yy,Haidenbauer:2005zh}. In contrast, our results indicate that the contribution from the $\Sigma N$ channel is relatively small and does not significantly influence the total cross section, except for a slight peak near the threshold. This behavior is similar to that predicted by the J\"ulich potential~\cite{Haidenbauer:2005zh}.

\begin{figure}[h!]
  \centering
  \includegraphics[
    width=0.48\textwidth,       
    keepaspectratio,           
    trim=0 0 0 0,             
    clip
  ]{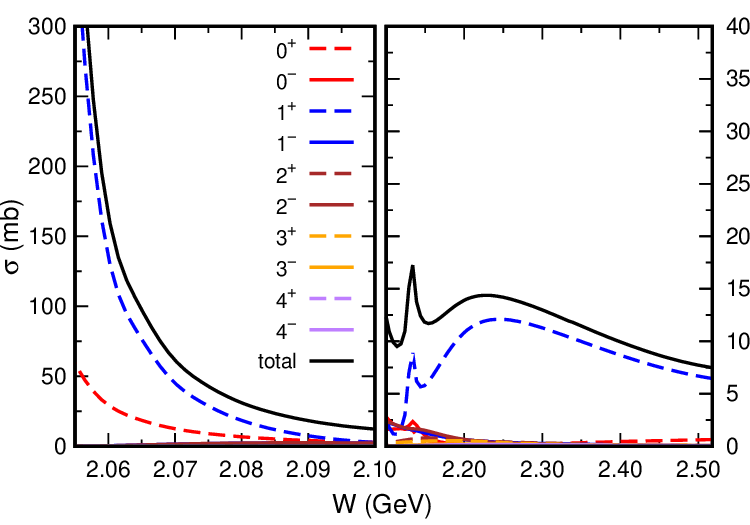}
\caption{Partial-wave cross sections for the scattering $\Lambda p \to \Lambda p$, showing the contributions from different spin-parity quantum numbers $J^P$. Results for energies below and above 2.1~GeV are presented in the left and right panels, respectively.}
  \label{fig:cs2}
\end{figure}

In our calculations, the cross sections are obtained by summing over the partial wave contributions with different spin-parity quantum numbers $J^P$. The results of the partial wave decomposition are shown in Fig.~\ref{fig:cs2}. It is found that the $1^+$ partial wave, represented by the blue dotted line, provides the dominant contribution to the total cross section of the $\Lambda p \to \Lambda p$ process. In the energy region from the $\Lambda p$ threshold up to 2.50 GeV, the $0^+$ partial wave also gives a significant contribution. As expected, the contributions from higher partial waves decrease rapidly with increasing energy. Furthermore, with increasing total angular momentum $J$, the contributions drop off quickly, and the partial waves with $J = 3$ and $J = 4$ are found to be very small, which ensures the convergence of the calculation up to $J = 4$.

\begin{figure}[!h] 
\centering
\includegraphics[width=0.52\textwidth,trim=0 20 0 20,clip]{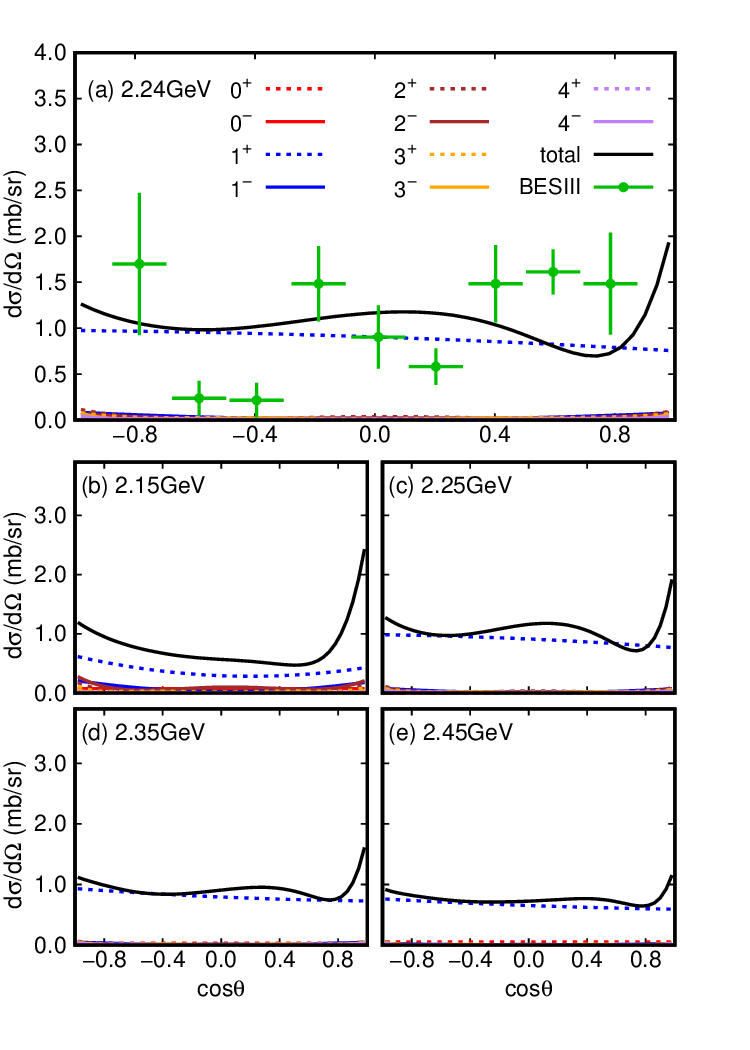}
\caption{Differential cross sections for the scattering process $\Lambda p \to \Lambda p$.  
Panel (a): Theoretical result at $\sqrt{s} = 2.24$~GeV, showing contributions from partial waves with different spin-parity $J^P$, compared with BESIII data (green filled circles)~\cite{BESIII:2024geh}.  
Panels (b)–(e): Predictions at $\sqrt{s} = 2.15$, 2.25, 2.35, and 2.45~GeV, respectively, also showing contributions from different $J^P$ partial waves.}
\label{fig:dif}
\end{figure}

The main result of the new BESIII observation is the measurement of the differential cross section for $\Lambda p$ scattering at $\sqrt{s} = 2.24$ GeV. Using the parameters fixed by fitting the total cross sections from threshold up to 2.5~GeV, we calculate the differential cross section at the same energy and compare it with the BESIII data, as shown in Fig.~\ref{fig:dif}(a). The theoretical result shows a relatively flat angular distribution with mild fluctuations. While these fluctuations are not entirely consistent with the experimental data, the overall behavior remains compatible within the sizable experimental uncertainties~\cite{BESIII:2024geh}.

We also present the contributions from individual partial waves with definite spin-parity $J^P$. Among them, the $1^+$ partial wave dominates. This $1^+$ component arises from a mixture of $S$- and $D$-waves. The $S$-wave part leads to a flat distribution, while the $D$-wave contribution distorts the flatness, resulting in a nontrivial shape for the $1^+$ component. Contributions from other partial waves are relatively small. Nevertheless, although their effects are negligible at intermediate angles, constructive interference among them enhances both forward and backward scattering slightly, leading to an overall tendency for forward peaking.

In addition, we provide predictions for the angular distributions at $\sqrt{s} = 2.15$, 2.25, 2.35, and 2.45 GeV, as displayed in Fig.~\ref{fig:dif}(b–e). These predictions consistently show a small forward enhancement and a comparatively weaker backward scattering behavior.

\subsection{Scattering $\bar{\Lambda} p \to \bar{\Lambda} p$}

\begin{figure}[h!]
  \centering
  \includegraphics[
    width=0.48\textwidth,  
    trim=0 0 0 0,     
    clip                  
    ]{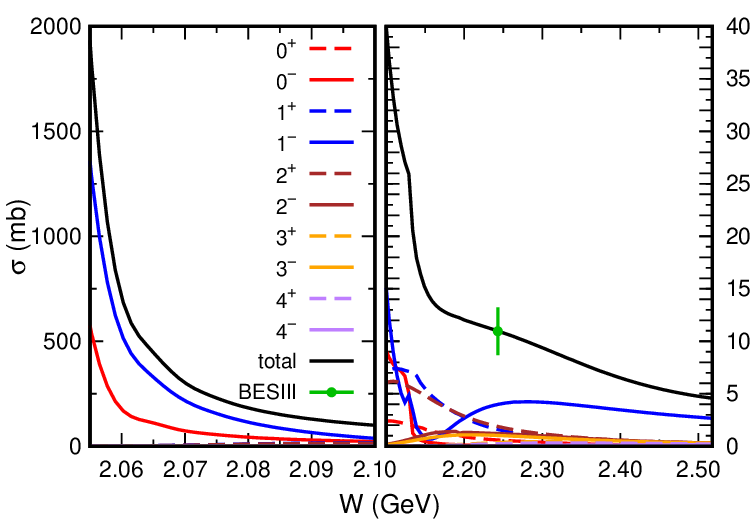}
\caption{Total cross section and partial-wave cross sections decomposed by spin-parity $J^P$ for the reaction $\bar\Lambda p \to \bar\Lambda p$. Experimental data (green filled circles) are taken from Ref.~\cite{BESIII:2024geh}.}
  \label{fig:csbar}
\end{figure}

Besides the differential cross section for $\Lambda p \to \Lambda p$ scattering, BESIII also provides results for the differential cross section of $\bar{\Lambda}p \to \bar{\Lambda}p$ scattering~\cite{BESIII:2024geh}. Although the $\Lambda p \to \Lambda p$ process has been extensively studied, as shown in Fig.~\ref{fig:cs1}, investigations of $\bar{\Lambda}p \to \bar{\Lambda}p$ remain relatively scarce. Since the $\bar{\Lambda}p \to \bar{\Lambda}p$ interaction shares most parameters with that of $\Lambda p \to \Lambda p$, we employ the model established above and fit the total cross sections of $\bar{\Lambda}p \to \bar{\Lambda}p$ in the energy range up to $2.500$ GeV using a slightly larger cutoff, $\Lambda = 0.69$ GeV. The resulting total and partial-wave cross sections are presented in Fig.~\ref{fig:csbar}.

As shown in Fig.~\ref{fig:csbar}, the total cross section for the $\bar{\Lambda} p \to \bar{\Lambda} p$ process exhibits a monotonically decreasing behavior in the low-energy region, from threshold up to approximately 2.1 GeV. The overall magnitude of the cross section is significantly larger than that of the corresponding $\Lambda p \to \Lambda p$ scattering process. Unlike the latter, however, no experimental data are currently available for $\bar{\Lambda} p \to \bar{\Lambda} p$ scattering, except for a recent measurement by the BESIII Collaboration at $\sqrt{s} = {2.24}$~GeV. Our model predicts a total cross section of $\sigma = 24.7$ mb at this energy, which is in good agreement with the BESIII result~\cite{BESIII:2024geh}.

We further analyze the partial-wave contributions to the total cross section, which reveal that the $1^-$ partial wave dominates over a wide energy range. In the region from threshold to 2.10 GeV, the $0^-$ wave also provides a non-negligible contribution. Between $\sqrt{s} = 2.10$ and 2.20 GeV, the $1^-$ partial wave shows a decreasing trend and ceases to be the dominant component. However, it regains dominance at energies above 2.25 GeV. The coupled-channel effects arising from the $\bar{\Sigma} N$ channel are included in our calculation, but their impact on the total cross section is found to be relatively minor.

\begin{figure}[!h] 
\includegraphics[width=0.52\textwidth,trim=0 20 0 20,clip]{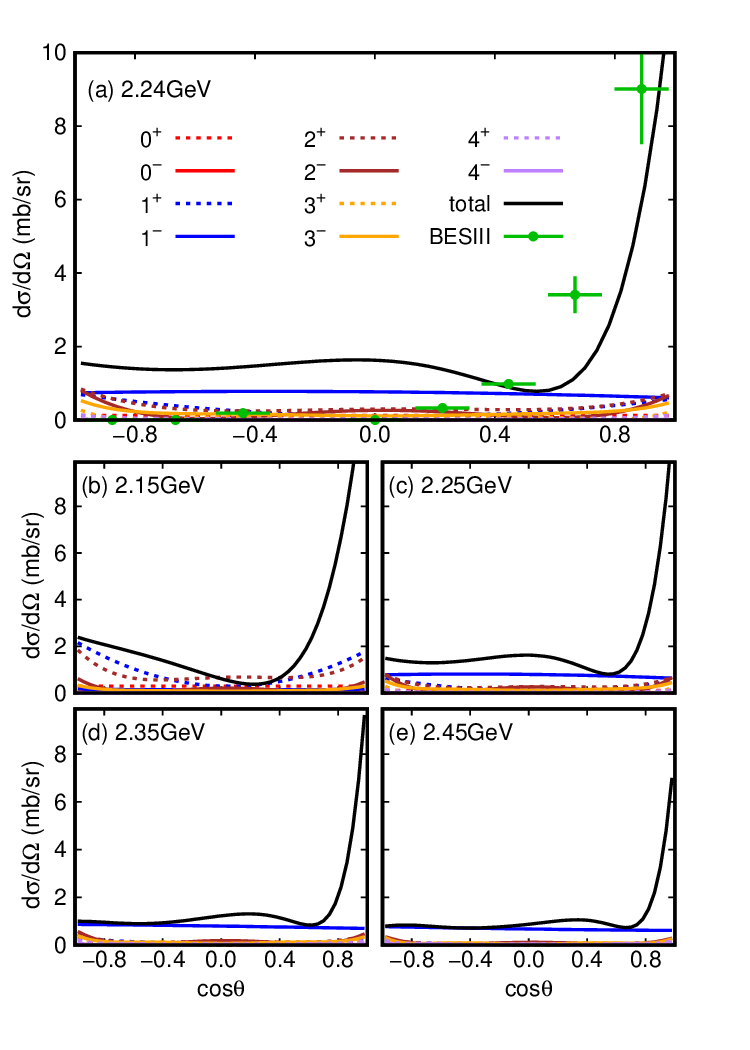}
\caption{Differential cross sections for the scattering process $\bar{\Lambda} p \to \bar{\Lambda} p$.  
Panel (a): Theoretical result at $\sqrt{s} = 2.24$~GeV, showing contributions from partial waves with different spin-parity $J^P$, compared with BESIII data (green filled circles)~\cite{BESIII:2024geh}.  
Panels (b)–(e): Predictions at $\sqrt{s} = 2.15$, 2.25, 2.35, and 2.45~GeV, respectively, also showing contributions from different $J^P$ partial waves.}
\label{fig:difbar}
\end{figure}

The differential cross sections for the $\bar{\Lambda}p \to \bar{\Lambda}p$ reaction at various center-of-mass energies are shown in Fig.~\ref{fig:difbar}. At $\sqrt{s} = 2.24$ GeV, a pronounced forward peak emerges in the angular distribution, exhibiting notable agreement with the BESIII data~\cite{BESIII:2024geh}. While the $1^-$ partial wave plays a prominent role, contributions from other partial waves remain non-negligible, particularly in the forward and backward regions. As discussed in Ref.~\cite{Wang:2024whi}, the absence of the $u$-channel mechanism can suppress backward enhancements, leading to a dominant forward structure. We infer that constructive interference among different partial waves at forward angles gives rise to the observed enhancement. This forward-peaking behavior persists at $\sqrt{s} = 2.15$, 2.35, and 2.45 GeV, with the peak becoming increasingly forward-focused as the energy increases.

\section{Summary}\label{sum}

In this work, we investigate the $\Lambda p$ and $\bar{\Lambda} p$ scattering
processes within the framework of the 
qBSE combined with an effective Lagrangian approach. The interaction
potentials are constructed using a one-boson-exchange model that includes
pseudoscalar, scalar, and vector meson exchanges. Coupled-channel effects from
$\Lambda p$--$\Sigma N$ and $\bar{\Lambda} p$--$\bar{\Sigma} N$ transitions are
also incorporated. The model parameters are constrained by fitting existing
total cross-section data, and the resulting framework is applied to calculate
differential cross sections, which are then compared with the BESIII
experimental results.

For the $\Lambda p \to \Lambda p$ scattering, we emphasize that our calculated total cross sections agree well with both experimental data and existing theoretical models at low energies. At higher energies, the model remains consistent with the experimental data and successfully reproduces the total cross section measured by BESIII. A mild enhancement appears near the $\Sigma N$ threshold, which can be attributed to coupled-channel effects. Partial wave analysis shows that the $1^+$ wave dominates throughout the studied energy range, while the $0^+$ contribution becomes noticeable near the threshold. Furthermore, the predicted angular distributions exhibit only a weak dependence on the scattering angle, in line with experimental observations.

For the $\bar{\Lambda}p \to \bar{\Lambda}p$ scattering, the calculated total cross section demonstrates a monotonic decrease with increasing center-of-mass energy, yielding a value of $\sigma = 24.7$ mb at $\sqrt{s} = 2.24$ GeV. This result is consistent with the recent BESIII measurement~\cite{BESIII:2024geh}. A partial wave decomposition reveals that the $1^-$ partial wave dominates the cross section at higher energies, while the $0^-$ wave remains non-negligible in the near-threshold region. The differential cross sections exhibit a pronounced forward peak over the entire energy range considered, in agreement with the experimental observations. This strong forward enhancement is attributed to constructive interference mechanisms that are particularly effective in the forward scattering region.

Overall, the present study provides a comprehensive analysis of $\Lambda p$ and $\bar\Lambda p$ scattering, offering new theoretical insights that are compatible with current experimental data. These results contribute valuable constraints on hyperon-nucleon interactions and offer guidance for future experimental investigations.

\section*{Acknowledgments}

This project is  supported by the National
Natural Science Foundation of China (Grant No.12475080).

\end{document}